\documentclass[aps,preprint,showpacs]{revtex4}

\usepackage[pdftex]{graphicx}
\usepackage{latexsym}
\usepackage{amsbsy}
\usepackage{amsmath}
\usepackage{amssymb}

\begin{document}

\date{\today}

\title{Manipulation of magnetic Skyrmions with a Scanning Tunneling
  Microscope}   

\author{R.\ Wieser}
\affiliation{1. International Center for Quantum Materials, Peking
  University, Beijing 100871, China \\  
2. Collaborative Innovation Center of Quantum Matter, Beijing 100871, China}

\begin{abstract}
The dynamics of a single magnetic Skyrmion in an atomic spin system
under the influence of Scanning Tunneling Microscope is investigated
by computer simulations solving the Landau-Lifshitz-Gilbert
equation. Two possible scenarios are described: manipulation with
aid of a spin-polarized tunneling current and by an electric field
created by the scanning tunneling microscope. The dynamics during the
creation and annihilation process is studied and the possibility to
move single Skyrmions is showed. 
\end{abstract}

\maketitle

Magnetic Skyrmions have been intensively studied during the last
period of time due to the possibility to use them as potential
candidates for data storage
\cite{zhangSR15III,zhouNatComm14,fertNatNano13}, for logic devices
\cite{zhangSR15}, or as Skyrmion transistor \cite{zhangSR15II}. The
idea to use local changes in the magnetic structure is not new: Bubble
domains in thin film structures \cite{odellRPP86}, magnetic domain
walls in nanowires driven by an electric current
\cite{parkinSCIENCE08}, and magnetic vortices 
\cite{wieserPRB06,yamadaNatureMat07ETAL} have been considered as
candidates for data storage and / or logic devices. However, due to
the stability as result of their topology (topological protected) and
their dimension (just a view nanometer) magnetic Skyrmions are
promising candidates for future spintronic devices. 

Furthermore, magnetic Skyrmions can be found in thin film systems on the
microscopic \cite{jiangSCIENCE15,wooNatMat16,muhlbauerSCIENCE09} but
also on the atomic length scale
\cite{rommingScience13,barkerPRL16}. In the moment  
most of the focus lies on the magnetic Skyrmions at the microscopic
length scale. The reason can be found in the possibility to observe
these Skyrmions at room temperature and with several experimental
techniques like e.g. magnetic transmission X-ray microscopy 
\cite{wooNatMat16} or magneto-optical Kerr effect microscopy
\cite{jiangSCIENCE15,linPRB16}. For Skyrmions at the atomic length
scale this is not the case. Here, a scanning tunneling microscope and
low temperatures ($T\approx 4$ K) are necessary. On the  
other hand the technological goal is to reduce the size which is
needed to store information. Therefore, magnetic Skyrmions on the
atomic length scale can be seen as the next step after realizing
Skyrmion devices on the microscopic scale.

The first step toward this direction has been already done. Recent
experiments have shown that it is possible to switch magnetic 
Skyrmions on the atomic length scale with aid of a scanning tunneling
microscope (STM) \cite{rommingScience13,hsuARXIV16}. Within these
publications two different methods to manipulate the Skyrmion have
been demonstrated: 1. creating and annihilating Skyrmions with
spin-polarized tunnel currents and 2. using electric fields. While for
the first experiment a magnetized STM tip is needed this is not the
case for the switching process using electric fields. Therefore, the
manipulation using electric fields is preferable.  
However, the problematic point in both experimental situations is the
time resolution. The time resolution of a conventional scanning
tunneling microscope is not high enough to investigate the dynamics
during the experiment. Therefore, spin dynamics simulations are
a perfect way to investigate the dynamics and to get deeper insights
in the physics of Skyrmions. 

The theoretical publications
\cite{hagemeisterNatComm14ETAL,siemensNJP16ETAL} discuss the ground state
configuration, the energy barrier between the ferromagnetic and the
Skyrmion state and give an idea about the dynamics. However, these
publications don't describe the creation or annihilation processes
using spin-polarized tunneling currents or electric fields. These
informations will be given in this Letter which provides a description of
the switching dynamics using STM in both cases spin-polarized with
tunneling currents and not spin-polarized using electric
fields. Furthermore, the possibility to move a single Skyrmion with
aid of an STM and without disturbing surrounding Skyrmions will be
demonstrated.  
 
The Manuscript is organized as follow: After introducing the investigated
system as well as the corresponding equation of motion the
Landau-Lifshitz-Gilbert equation with additional spin torque terms,
the dynamics during the creation and annihilation process will be
described and the possibility to move a single Skyrmion with an STM
demonstrated. Thereby, both experimental ways: the
usage of a spin-polarized tunneling current as well as the an electric
field are considered. The Letter ends with a summary.       

The investigated system is a spin system on the atomic length scale with 
a triangular lattice similar to the double layer Pd/Fe on Ir(111)
described in \cite{hagemeisterNatComm14ETAL}. The lateral dimension of
the film is $L_x \times L_y = 26.325$ nm $\times$ 45.6 nm (39204
spins) and the magnetic properties of the system are well described by
the following Hamiltonian: 
\begin{small}   
\begin{eqnarray} \label{Ham}
{\cal H} = -J \sum\limits_{\langle n,m \rangle} \mathbf{S}_n \cdot \mathbf{S}_m 
-  \sum\limits_{\langle n,m \rangle} {\boldsymbol{\cal D}}_{nm} \cdot
(\mathbf{S}_n \times \mathbf{S}_m) - \mu_S B_z \sum\limits_n S_n^z
\;. \nonumber
\end{eqnarray}
\end{small}
The first two terms are the ferromagnetic exchange and
Dzyaloshinky-Moriya interaction (DMI) where $J = 7$ mev and $|{\boldsymbol{\cal
D}}_{nm}| = 2.2$ meV have been assumed. The Dzyaloshinky-Moriya vectors
${\boldsymbol{\cal D}}_{nm}$ are oriented in-plane (film plane)
perpendicular to the lattice vector $\mathbf{r}_{nm} = \mathbf{r}_m -
\mathbf{r}_n$ pointing from lattice site $n$ to $m$. The third term
describes the influence of an external magnetic field perpendicular to
the film plane in $z$-direction. 

\begin{figure}
\vspace{1mm} 
\includegraphics*[width=7.cm,bb = 216 492 349 582]{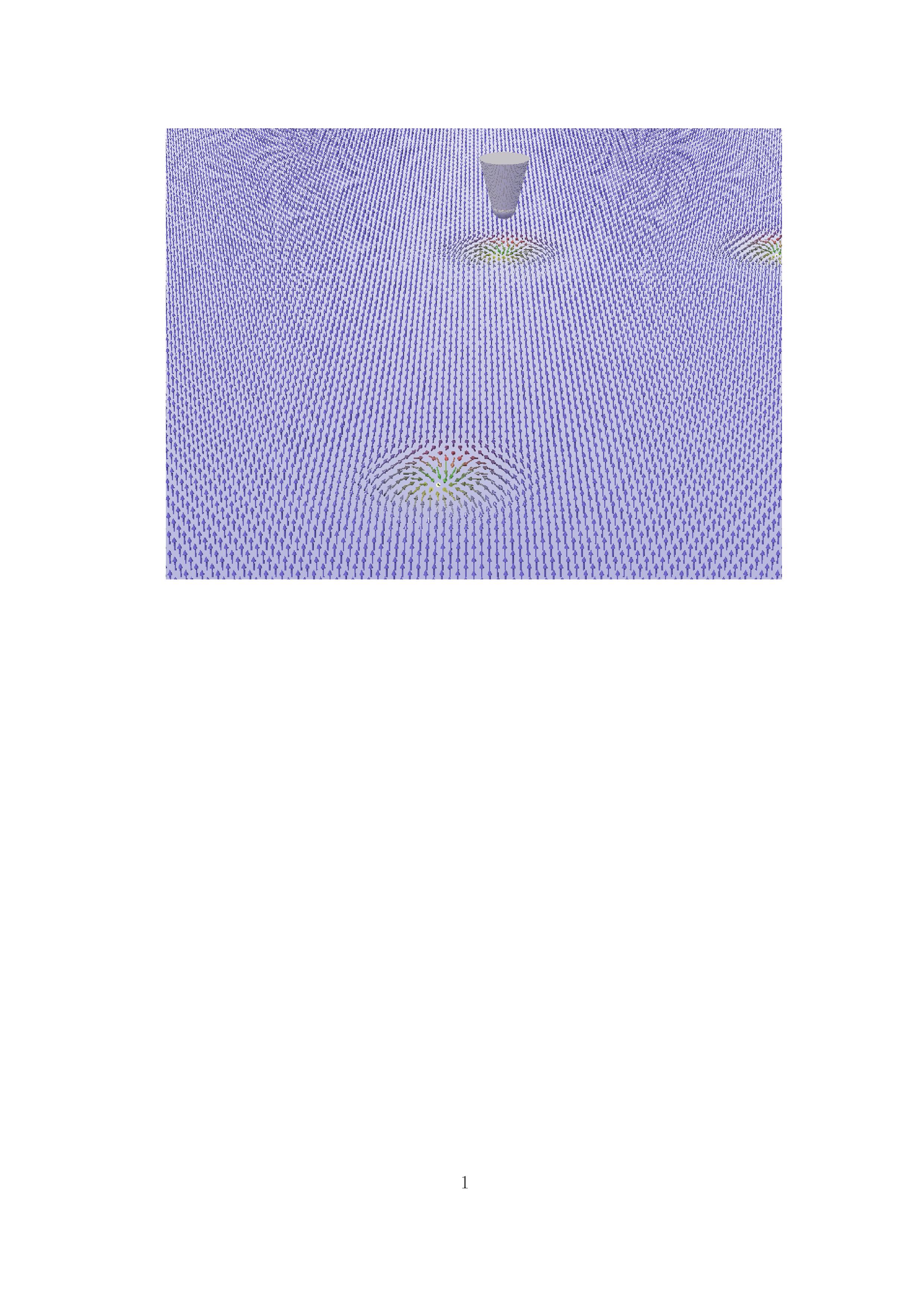}
\includegraphics*[width=7.cm,bb = 216 492 349 582]{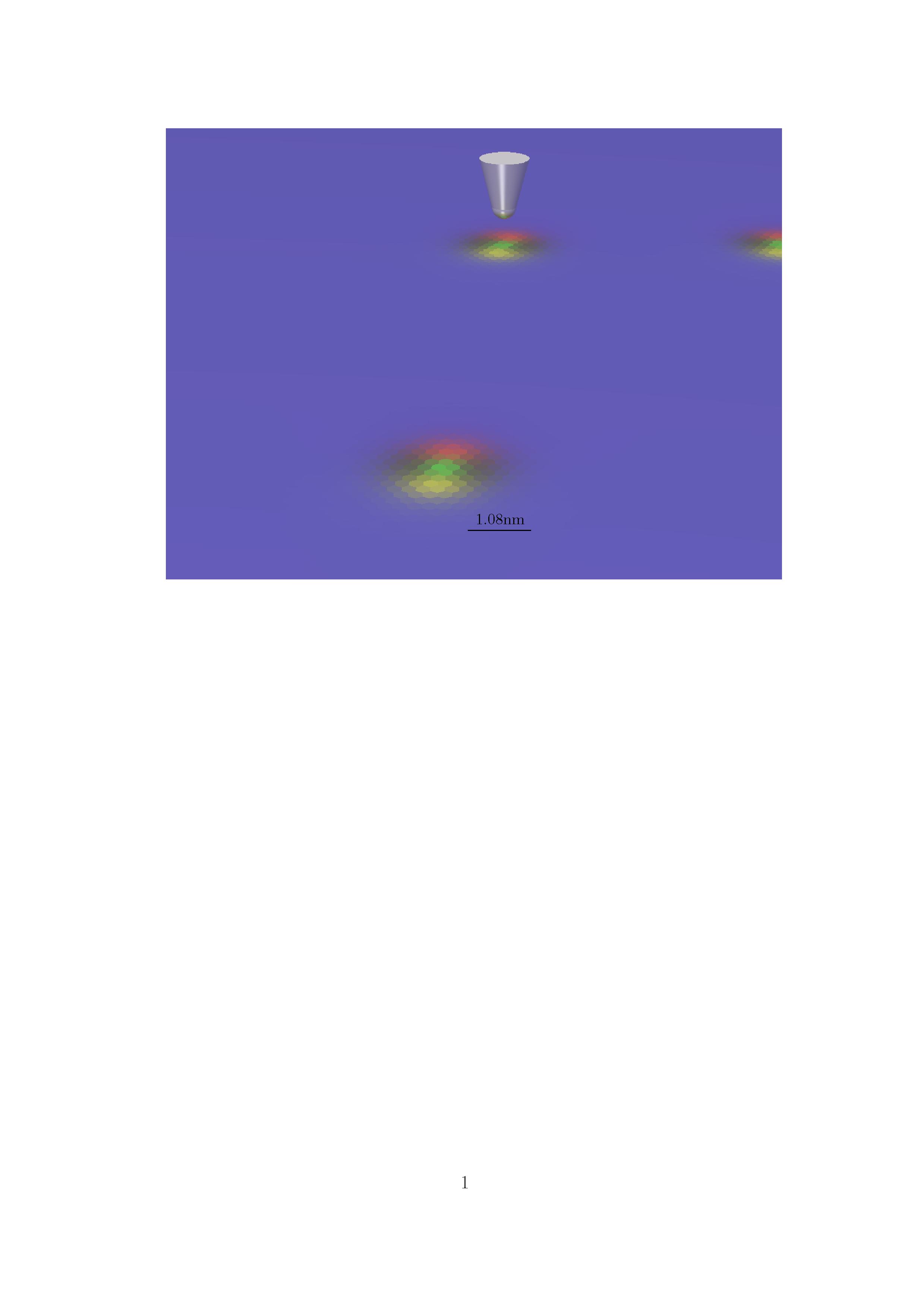}  
  \caption{(color online) Single Skyrmion in the Skyrmion gas phase
    \cite{ezawaPRB11} at $B_z = 7$ T. Left: triangular lattice hedgehog
    structure, right: corresponding continuum (color) picture. 
}       
  \label{f:pic1}
\end{figure}

Without magnetic field $B_z = 0$ T the system shows a maze like spin spiral
structure \cite{hagemeisterNatComm14ETAL}. With external field the  
magnetic configuration provides Merons \cite{ezawaPRB11} and at larger fields
($B_z$ approximately in between 4 and 10 T) Skyrmions 
which due to the in plane orientation of ${\boldsymbol{\cal D}}_{nm}$
show a hedgehog structure with magnetic moments pointing to the center
of the Skyrmion. For $B_z \geq 10$ T the ferromagnetic state is the
ground state and Skyrmions are no longer existent. The Skyrmionic
structure as well as the color coding of the pictures are given in
Fig.~\ref{f:pic1}. All figures have the same camera position and
therefore the same color coding even if the focus varies.
The diameter of the Skyrmion depends on the strength of the external
field. During the simulations $B_z = 7$ T (Skyrmion creation and
destruction) and $B_z = 4.5$ T (manipulation) have been
used. Therefore, the corresponding Skyrmion diameter are $d \approx
3.24$ nm ($B_z = 7$T) and $d \approx 10.26$ nm ($B_z = 4.5$T). 

The dynamics of the system is described by the Landau-Lifshitz-Gilbert (LLG)
equation with additional spin transfer torques:
\begin{eqnarray}
 \label{LLG}
 \frac{\partial {\mathbf S}_n}{\partial t} = &-&
\frac{\gamma}{\left(1+\alpha^2\right)\mu_S}{\mathbf S}_n \times \left[
  {\mathbf H}_n + \alpha \left({\mathbf S}_n \times {\mathbf H}_n \right)\right]
  \nonumber \\  
&-& {\mathbf S}_n \times \left[\mathcal{A} {\mathbf T}_n + \mathcal{B}
    \left({\mathbf S}_n \times {\mathbf T}_n \right)\right]\;. \nonumber
 \end{eqnarray}
The first and second term are the precessional and
relaxation term of the conventional LLG equation with the effective
field: ${\mathbf H}_n = -\partial{\cal H}/\partial{\mathbf S}_n +
{\boldsymbol \xi}_n$, where ${\boldsymbol \xi}_n$ is a white noise which
simulates the effect of temperature. For the description of the
manipulation via electric fields underneath the tip (radius $r \approx 0.6$
nm) a locally increased temperature: $7.5$ K (Joule heating) has been taken
into account. During the simulations describing the manipulation using
spin-polarized tunnel currents no temperature effects have been taken
into account. The other parameters in the LLG equation are the
gyromagnetic ratio $\gamma = 1.76 \cdot 10^{11}$ $\frac{1}{Ts}$, the
magnetic moment $\mu_S = 3.3$ $\mu_B$ in Bohr magneton $\mu_B$, and
the dimensionless Gilbert damping constant $\alpha = 0.02$. The third
and fourth term are spin transfer torques describing the influence 
of an spin polarized current. These terms have been modified to
describe the tunnel current of a spin-polarized scanning tunneling
microscope: $\mathcal{A} = 0.05$ and $\mathcal{B} = 1.0$ have been
assumed. The model which describes the local strength of the
current is the Tersoff-Hamann model which leads to:   
\begin{equation}\label{Ttip}
 {\mathbf T}_n =
 -I_0e^{-2\kappa\sqrt{(x_n-x_{\mathrm{tip}})^2+(y_n-y_{\mathrm{tip}})^2+h^2}}{\mathbf
   P}\;, \nonumber
\end{equation}
where $\mathbf{P} = \pm \hat{\mathbf{z}}$ is the polarization of the tip,
$I_0$ the strength of the current, $\kappa = 0.93\cdot10^{10}$ 1/m is
related to the work function of the tip and $\mathbf{r}_{\mathrm{tip}} =
(x_{\mathrm{tip}},y_{\mathrm{tip}},h)^T$ is the tip position which is
time dependent, and $\mathbf{r}_n = (x_n,y_n,0)$ the position vector
within the lattice. 
\begin{figure}
\vspace{1mm} 
\includegraphics*[width=7.cm, bb = 170 635 365 760]{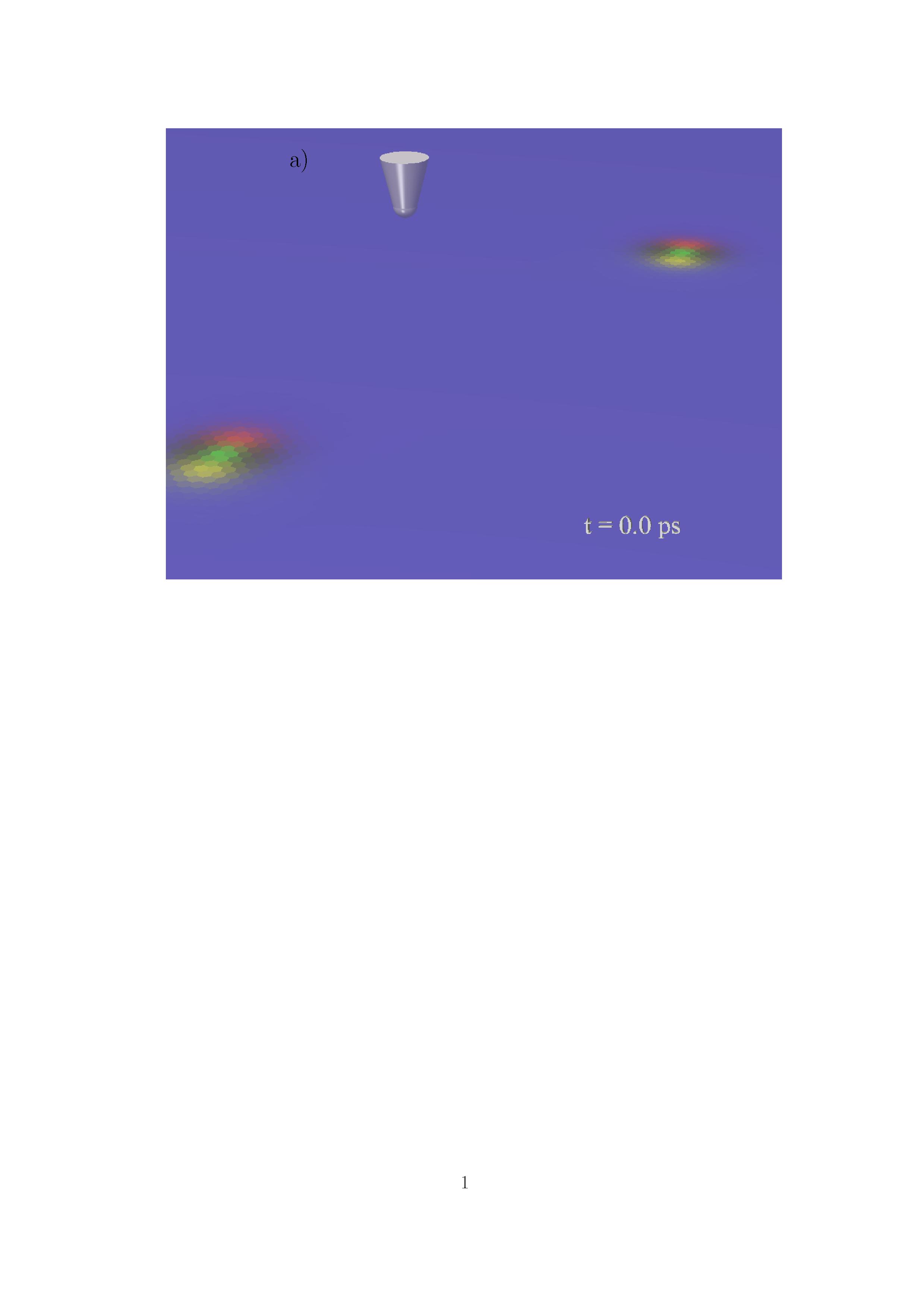}
\includegraphics*[width=7.cm, bb = 170 635 365 760]{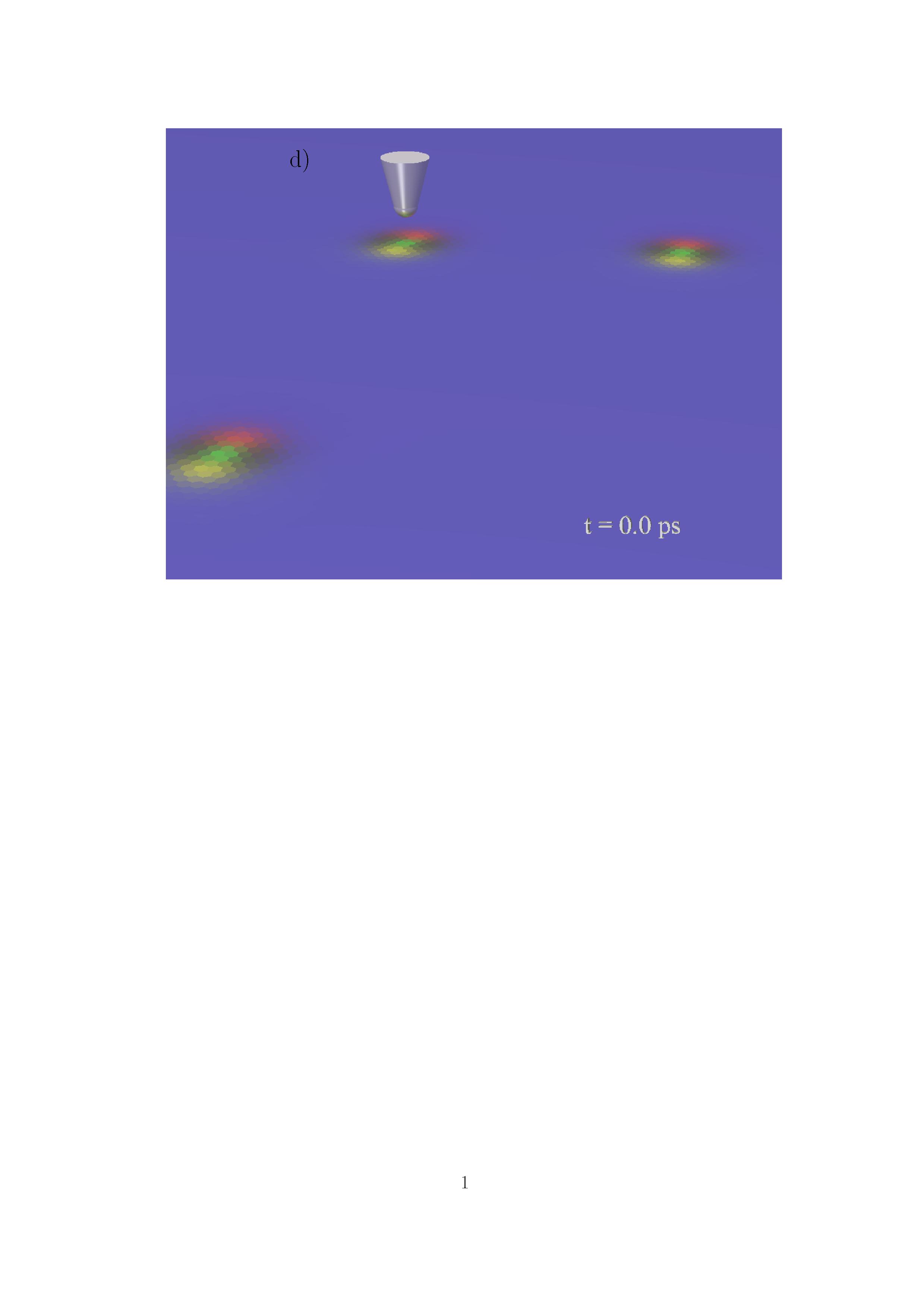}
\includegraphics*[width=7.cm, bb = 170 635 365 760]{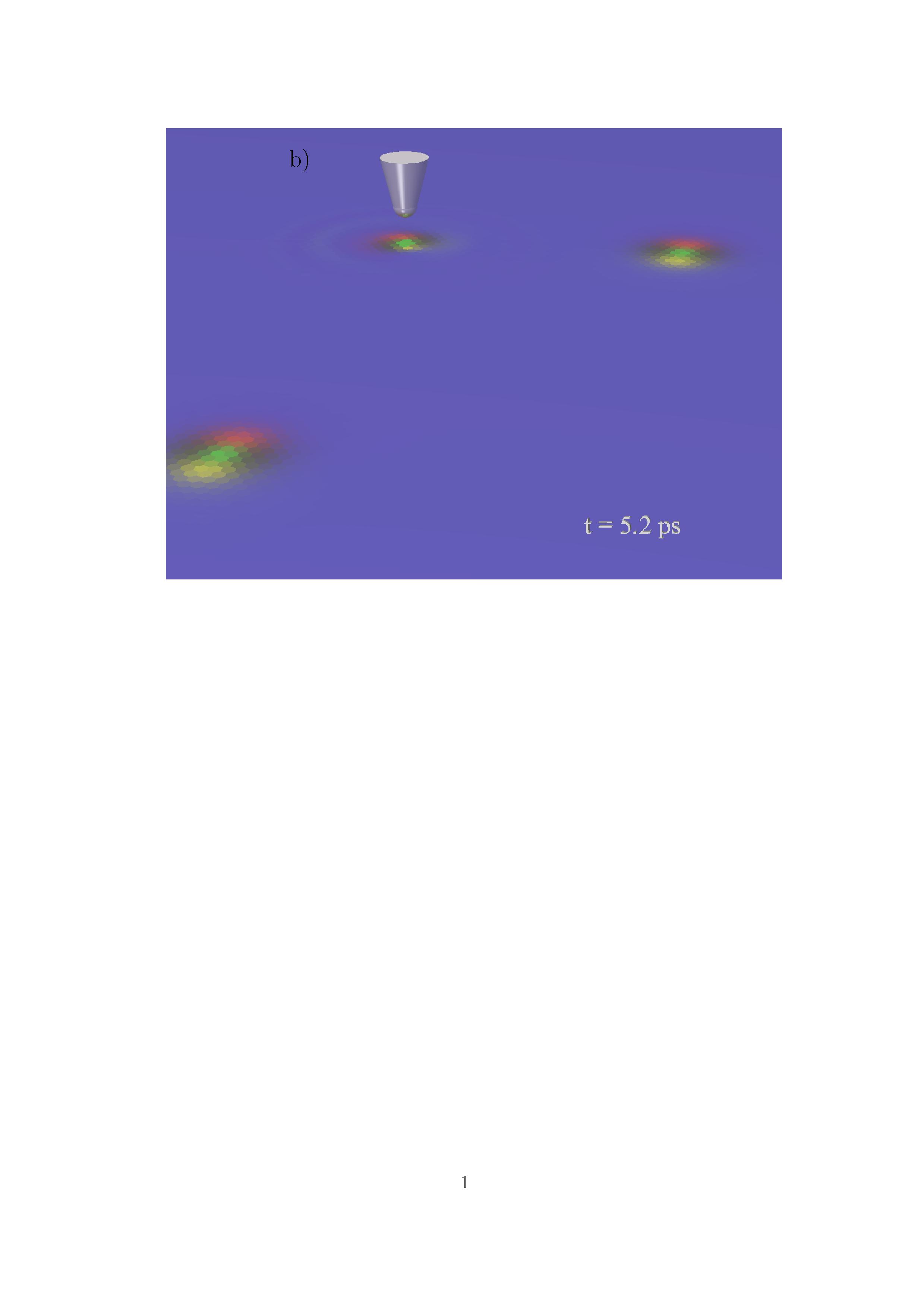}
\includegraphics*[width=7.cm, bb = 170 635 365 760]{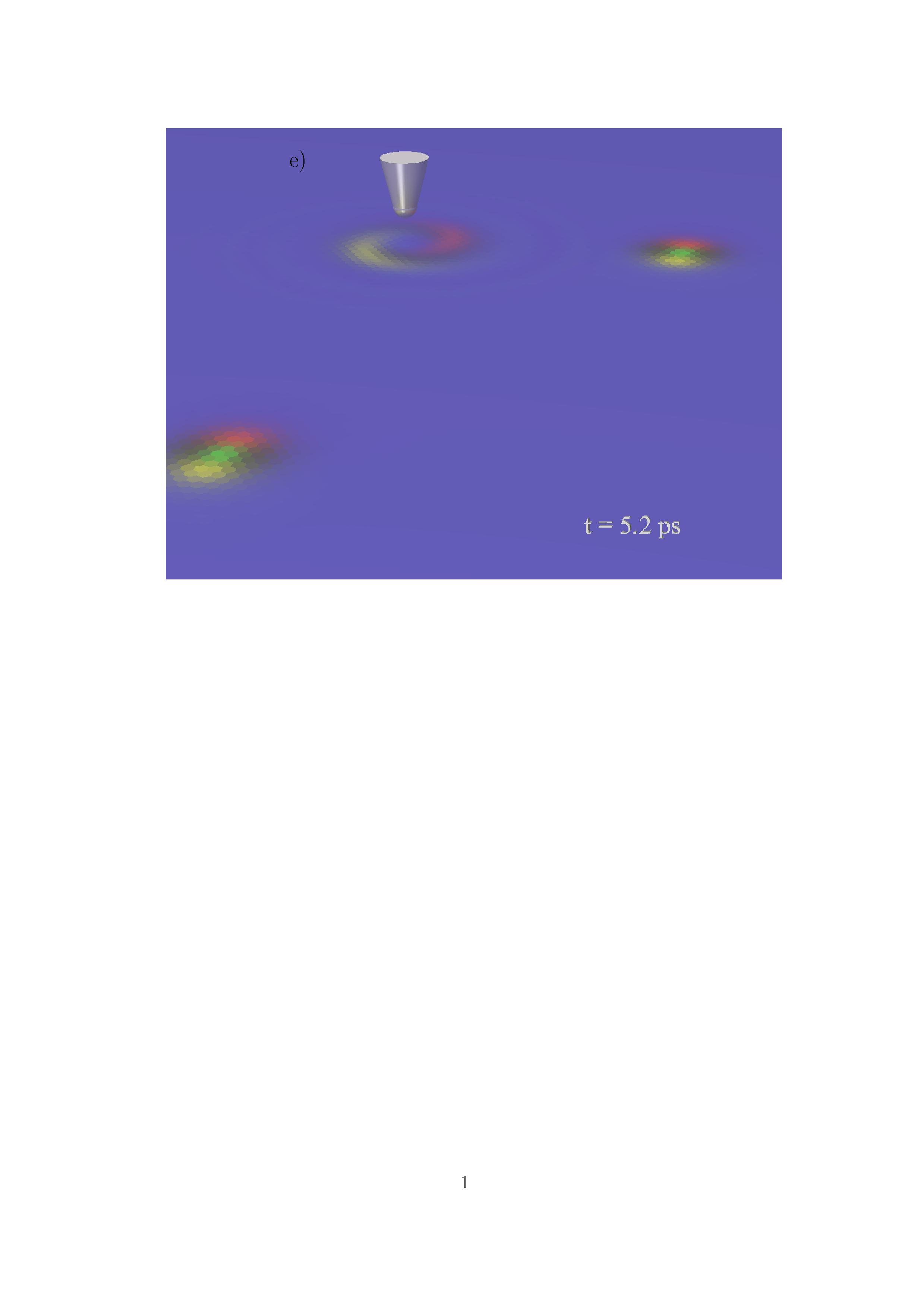} 
\includegraphics*[width=7.cm, bb = 170 635 365 760]{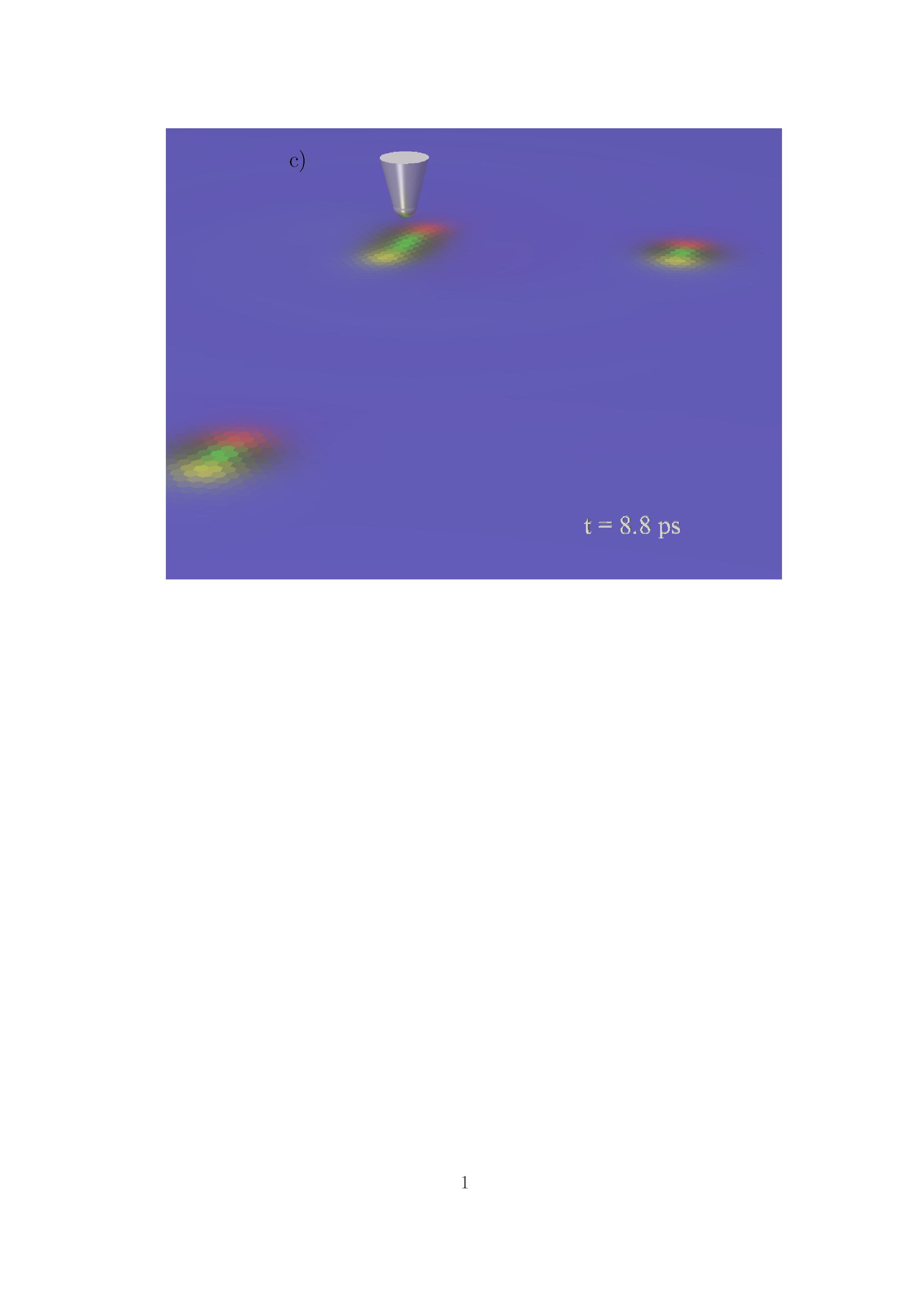}
\includegraphics*[width=7.cm, bb = 170 635 365 760]{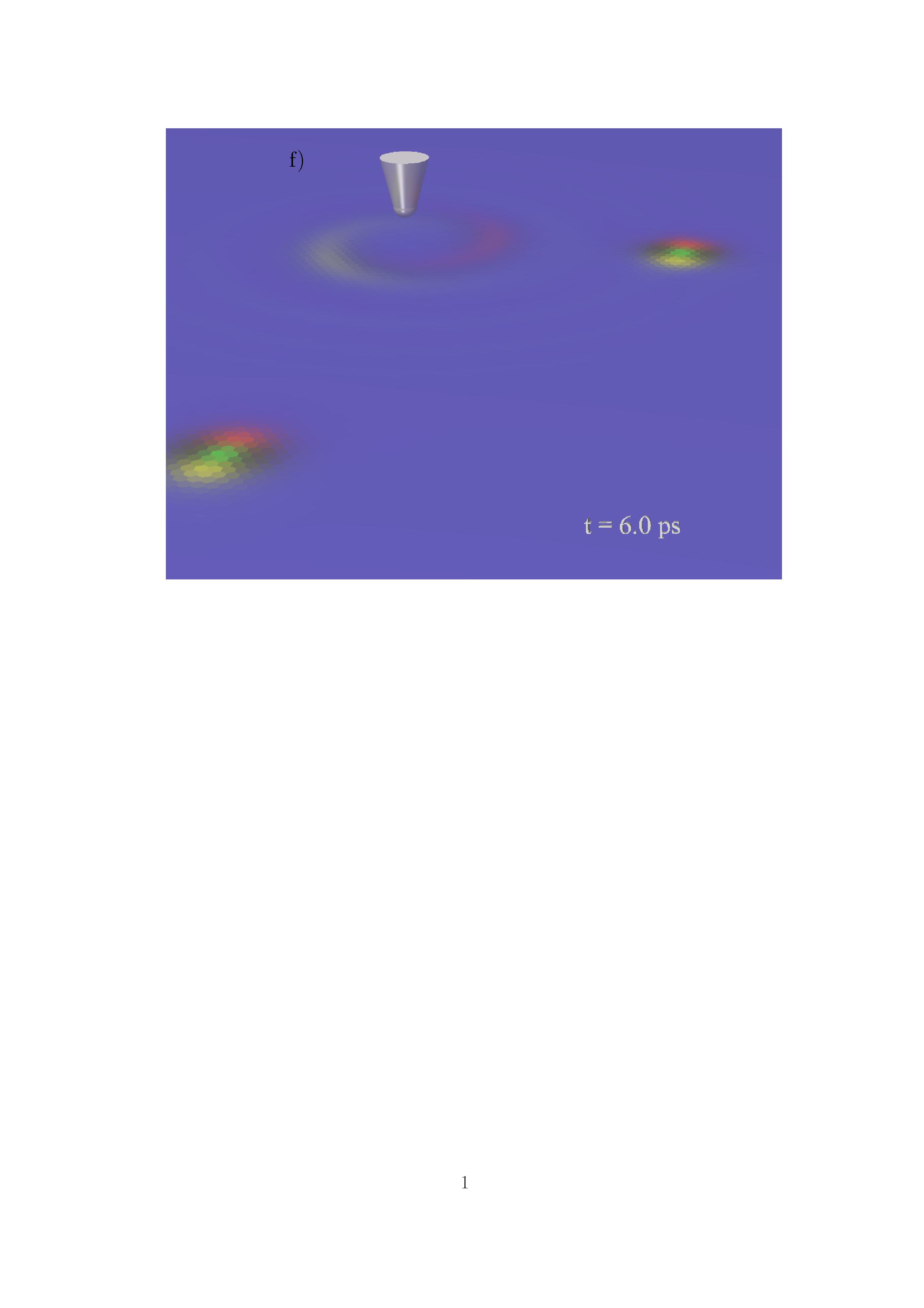}  
  \caption{(color online) Creation and annihilation of a Skyrmion with
    current pulses from a spin-polarized STM tip. Left column:
    creation with tip polarization $\mathbf{P} = -\hat{\mathbf{z}}$,
    right column: annihilation with tip polarization $\mathbf{P} =
    +\hat{\mathbf{z}}$.  
}       
  \label{f:pic2}
\end{figure}

\begin{figure*}
\vspace{1mm} 
\includegraphics*[width=8.cm, bb = 106 610 503 708]{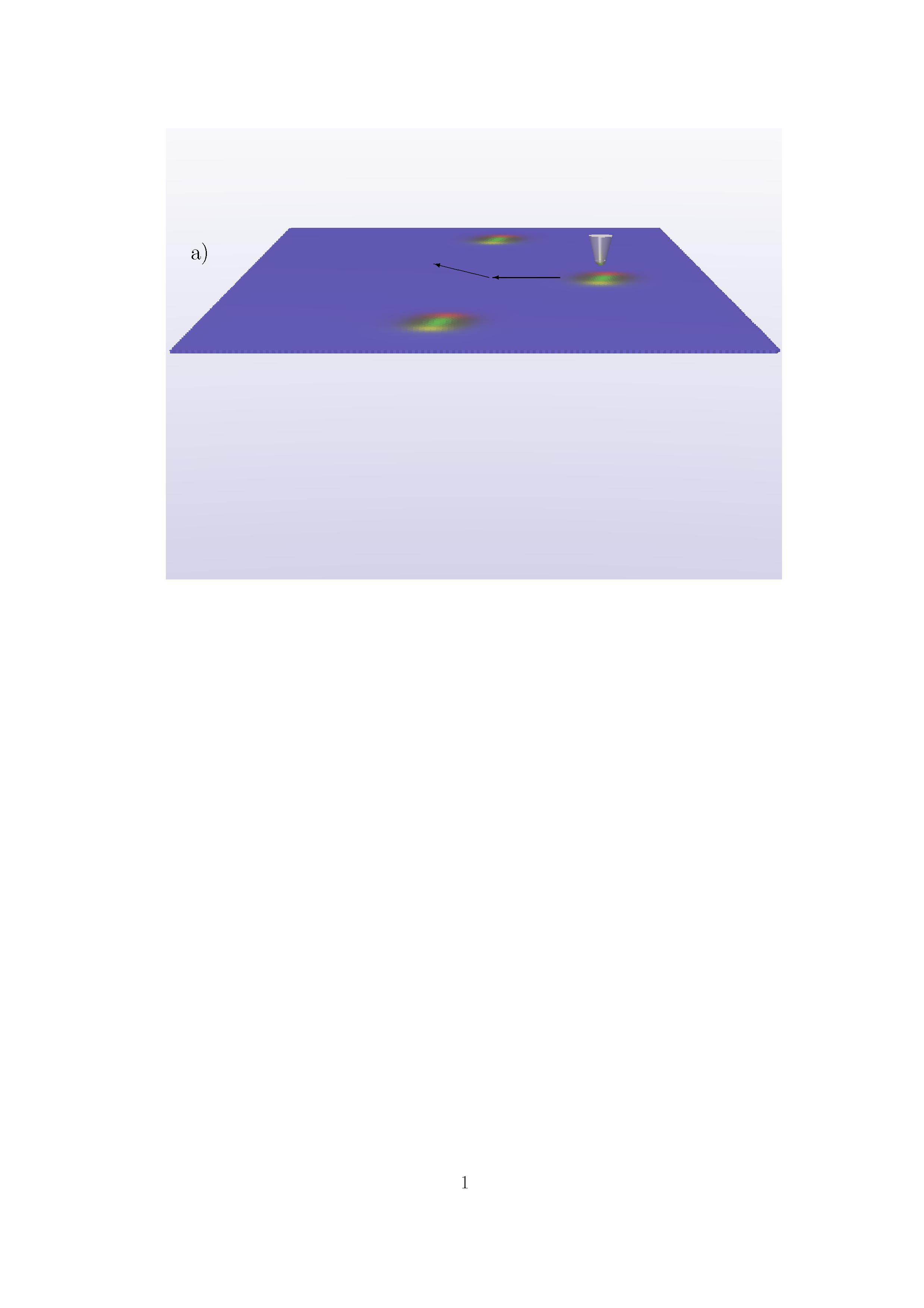} 
\includegraphics*[width=8.cm, bb = 106 610 503 708]{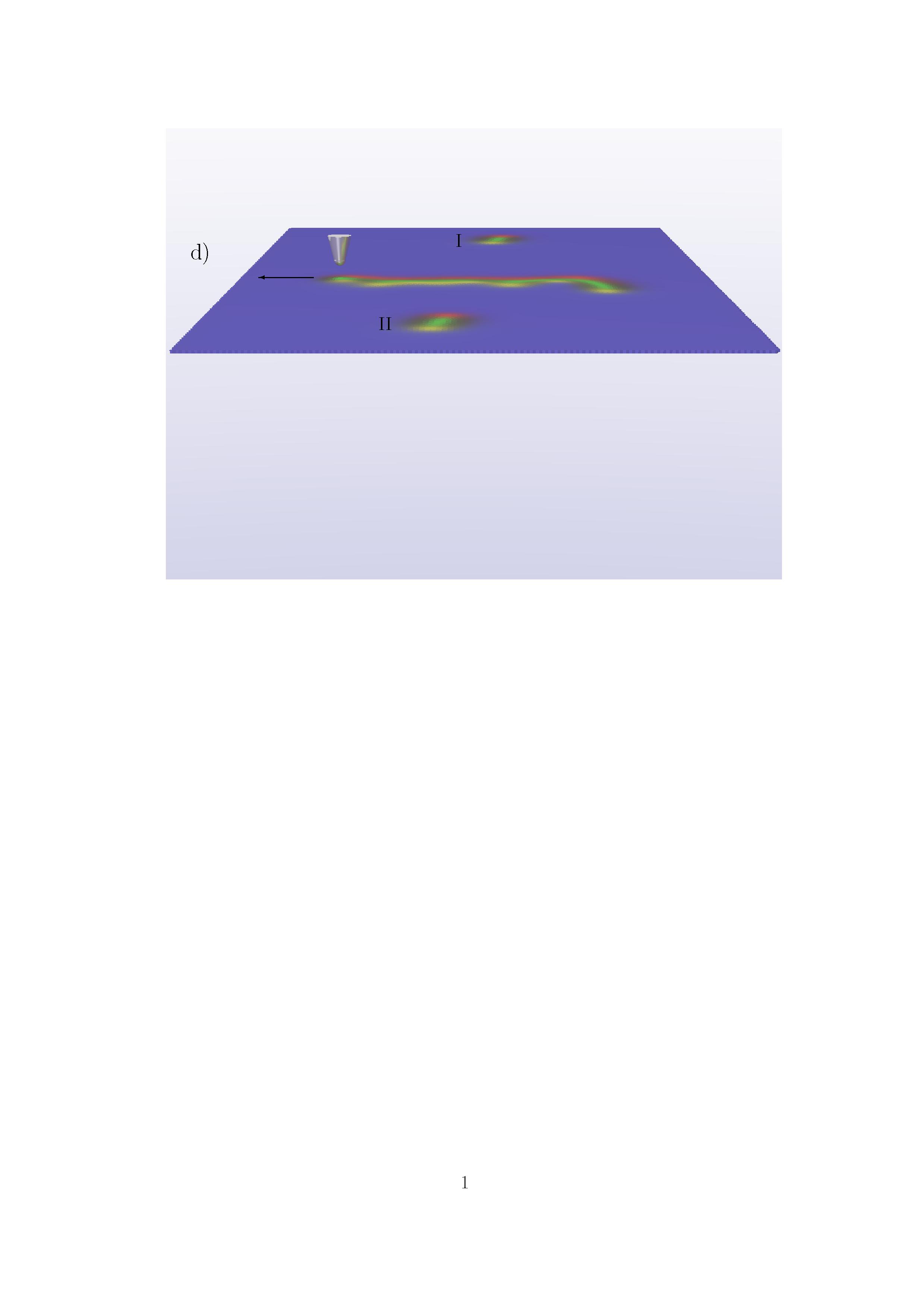} 
\includegraphics*[width=8.cm, bb = 106 610 503 708]{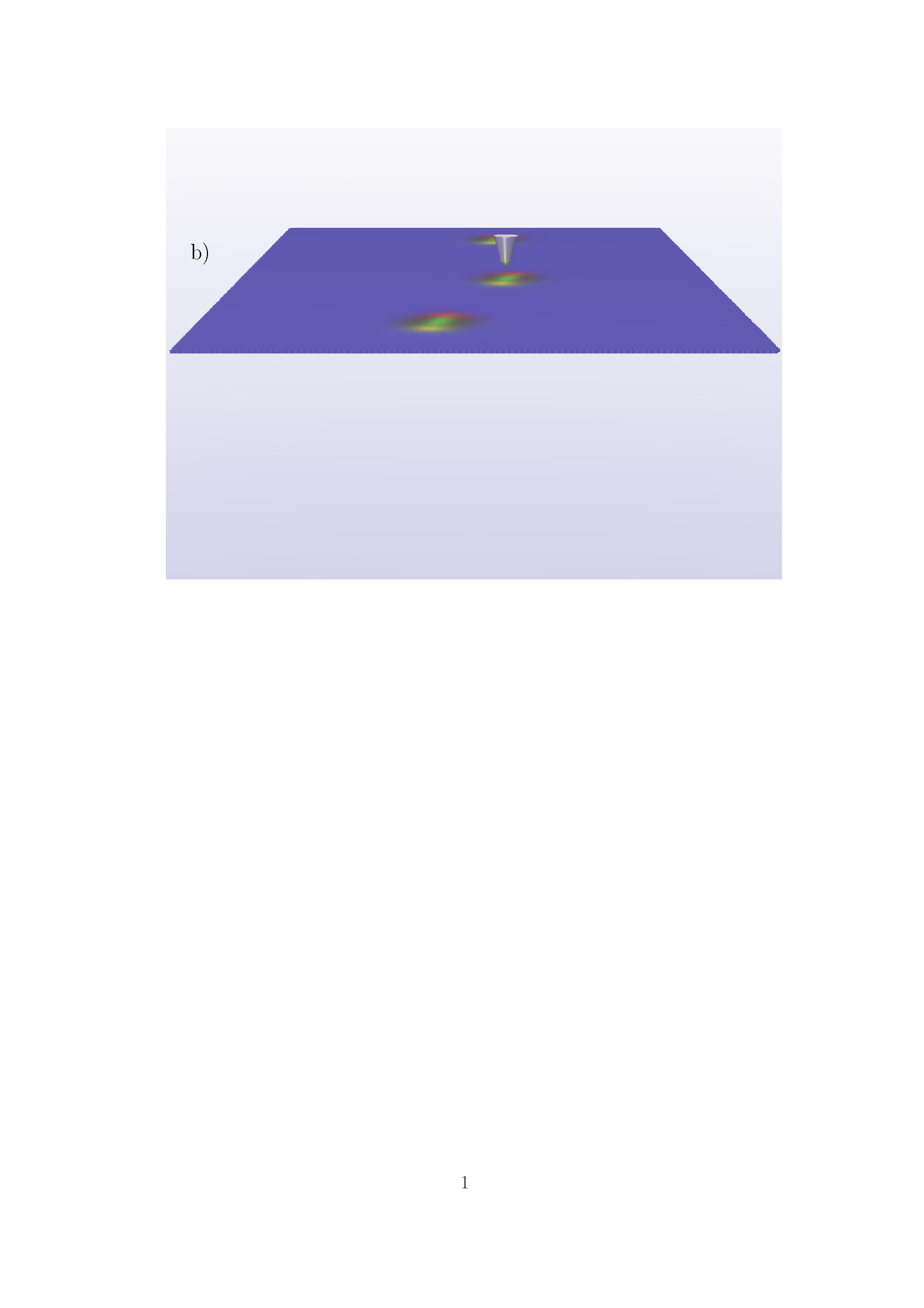} 
\includegraphics*[width=8.cm, bb = 106 610 503 708]{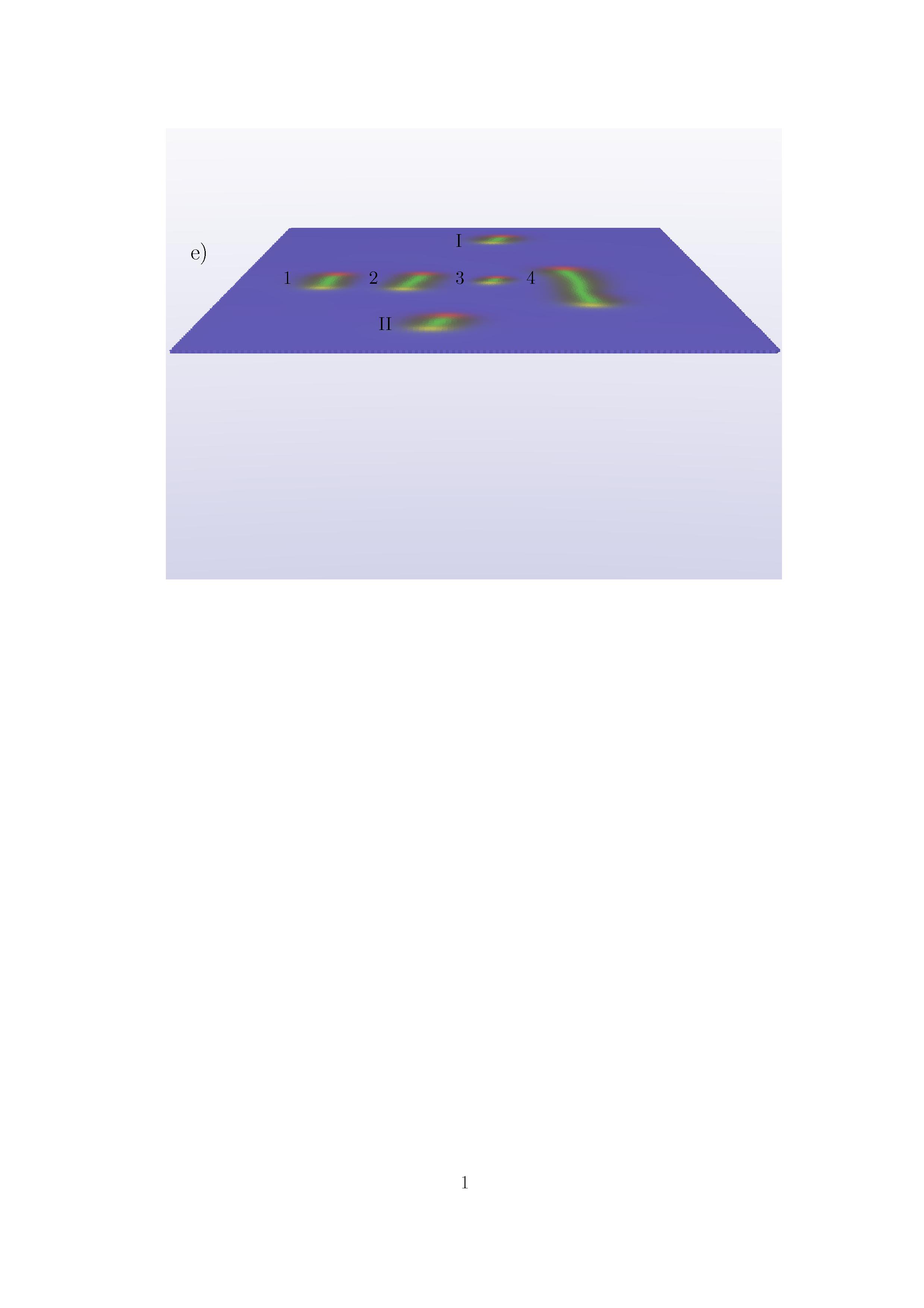} 
\includegraphics*[width=8.cm, bb = 106 610 503 708]{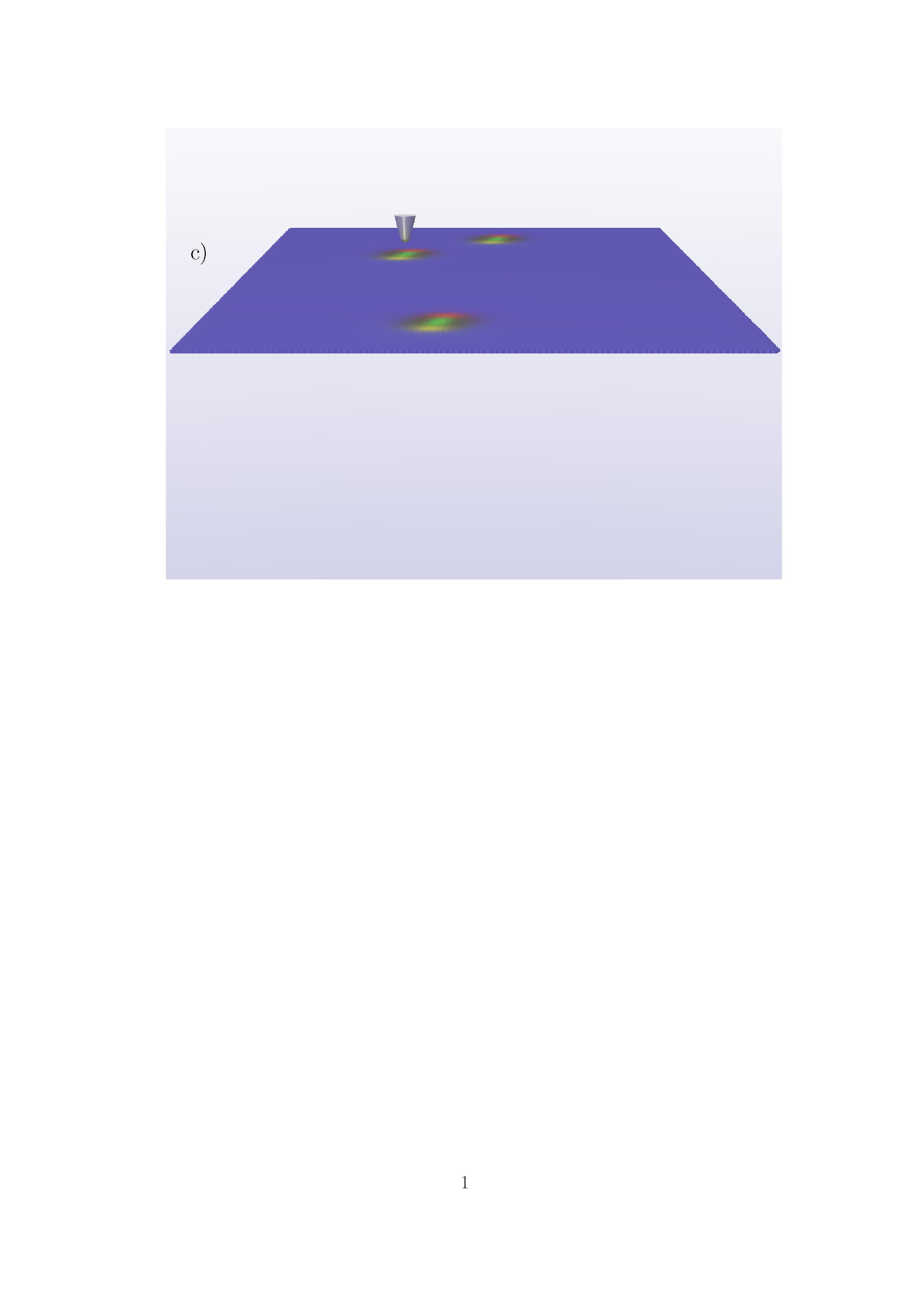} 
\includegraphics*[width=8.cm, bb = 106 610 503 708]{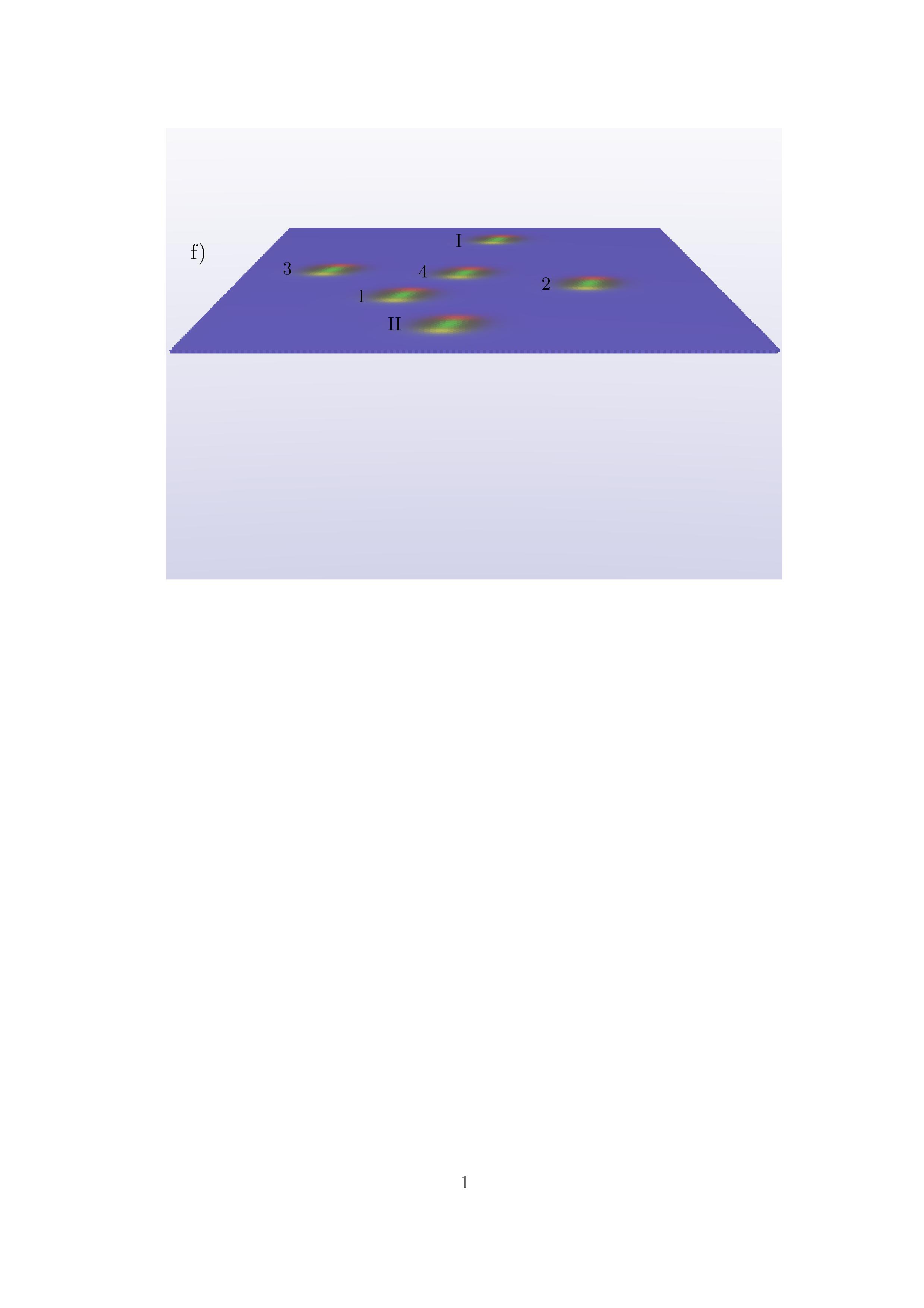} 
  \caption{(color online) Manipulation of a single Skyrmion with a
    spin-polarized STM. Left column: slow tip velocity and ``normal''
    tunnel current strength, right column: huge tip velocity and large
    tunnel current. The Roman and Arabic numerals give the positions
    of the original and new created Skyrmions. Due to the interaction
    a clear change in the order can be seen.     
}       
  \label{f:pic3}
\end{figure*}
Fig.~\ref{f:pic2} provides a sequence of pictures showing the
creation and annihilation of a magnetic Skyrmion with a spin-polarized
tunnel current. The starting configuration with a ferromagnetic
orientation of the magnetic moments underneath the STM tip is
given in Fig.~\ref{f:pic2}a). After a current pulse of $8$ ps with
$I_0^{\mathrm{max}} = 1.27\cdot 10^{19}$ 1/s a Skyrmion is 
created. In total the creating process takes just a few pico
seconds. Fig.~\ref{f:pic2}b) and c) show two moments during the   
creation of the Skyrmion. The process starts with the
reversal of the magnetic moments underneath the magnetic tip 
forced by the tunnel current. The tip polarization $\mathbf{P}$ is
assumed to be in $-z$-direction which is opposite to the orientation of the
ferromagnetically aligned magnetic moments. After a short moment the
Skyrmion configuration is created and the Skyrmion grows with the
time. The size of the Skyrmion overpasses the size of the final
configuration before it shrinks. During the shrinking the Skyrmion
shows first a nonlinear excitations: fuzzy oscillation which becomes a
breathing mode \cite{liuPRB15,kimPRB14} with frequency $f \approx
150$GHz. The final configuration is the Skyrmion shown in
Fig.~\ref{f:pic2}d). Another, current pulse with opposite tip
polarization can be used to annihilate the
Skyrmion. Fig.~\ref{f:pic2}e) and f) show two snap shots of the 
destruction process. While during the Skyrmion creation the tunneling
current has created a point singularity which becomes the center of
the Skyrmion in the case of the annihilation process the tunnel
current orients all spins underneath the tip parallel to the tip
polarization. This means the tunnel current destroys the singularity
and therefore the Skyrmion. It is a known fact that the central singularity is
responsible for the stability of the Skyrmion (topological
protection). Due to this singularity the Skyrmion has another
topological charge than the Ferromagnet and the Skyrmion which
stabilizes the Skyrmion. 

Due to the locally appearance of the reversal process (just underneath
the tip) and the fact that the system gains energy during the
destruction of the Skyrmion (more precisely the singularity) the
annihilation process appears like an explosion where a concentric shock
wave starting underneath the tip and exciting the surrounding Skyrmions can be
seen. The energy win just for the central spin can be easily
calculated. In the case of a collinear structure and under the
assumption of the Hamiltonian $\cal H$ a reversed central
spin with 6 nearest neighbors will set free an energy of $\Delta E =
6J + 2\mu_S B_z$.

Despite the creation and annihilation of Skyrmions the scanning
tunneling microscope can be also used to move Skyrmions. The
possibility to shift domain walls has been discussed in
\cite{stapelfeldtPRL11ETAL,wieserEPL12ETAL}. Fig.~\ref{f:pic3}a) shows a
sequence of pictures where the movement of a single Skyrmion with a
moving scanning tunneling tip is demonstrated. The tip has a
polarization in $-z$-direction which is opposite to the orientation of
the ferromagnetic surrounding of the Skyrmion but parallel to the
orientation in the center. In the former publications it has been shown
that a tip polarization parallel to the magnetization in the center of
the domain wall is the best choice. In the case of the Skyrmion a tip
polarization parallel to the central magnetization is the best decision
and has been used during the simulations. Furthermore, $I_0$ has been
set to $I_0 = 2.0\cdot 10^{17}$ 1/s. At this point it has to be said
that the velocity of the domain wall is equal to the velocity of the
tip. This can be seen as an disadvantage because the STM tip velocity is
very low. What happens if the tip moves to fast can be seen in
Fig.~\ref{f:pic3}b). Here, a larger current intensity ($I_0 = 2.0\cdot
10^{19}$ 1/s) has been assumed. In this case the current creates a
Meron which is not stable. The Meron collapses into several shorter
Merons which become under rotational movements Skyrmions. The
rotations of these Merons lead to interaction between them in such a
way that they push each other. During the simulation a rotation of two
Skyrmions around each other can be  seen. The Roman and Arabic numbers
in Fig.~\ref{f:pic3}d)-f) give the positions of the original and
created Skyrmions. This scene remembers the rotational motion of two
identical masses around the center of mass and has been reported in
the case of Skyrmions in multilayer systems \cite{daiSciRep14} as well
as for pairs of vortices \cite{voelkelPRB91,buchananNatPhys05}.       
 
In the previous investigations within this Letter the Skyrmion has
been manipulated with aid of a spin-polarized tunnel current. However,
using spin polarized current means that it is needed to have a stable
spin-polarized tip. Furthermore, a tip polarization is needed which is
opposite to the external field. Due to these conditions it is more
effective if the manipulation could be done with an unpolarized
tip. Recently Hsu et al. \cite{hsuARXIV16} have shown that it is
possible to create and annihilate Skyrmions by using electric field
pulses. The performed simulations show that all the scenarios
presented before: moving, creating and annihilating of Skyrmions can
be done with an electric field. The electric field itself is local and
affects locally the DMI \cite{siratoriJPSJ80,chenPRL15}:   
\begin{equation}
{\boldsymbol{\cal D}}_{nm} = {\boldsymbol{\cal D}}_{nm}^0 +
\omega_{nm} (\mathbf{E} \times \mathbf{r}_{nm}) \;. \nonumber
\end{equation} 
${\boldsymbol{\cal D}}_{nm}^0$ is the original DMI without electric
field, $\mathbf{E}$ is the electric field vector, $\mathbf{r}_{nm} =
\mathbf{r}_n - \mathbf{r}_m$ is the vector pointing from lattice site
$n$ to lattice site $m$, and $\omega_{nm}$ is a constant. If, as in the
experiment the electric field is given by a vector pointing in $\pm
z$-direction (perpendicular to the film plane) $\mathbf{E} \times
\mathbf{r}_{nm}$ will be parallel/antiparallel to the DMI vector
and increases or decreases the effect of the DMI. This means
especially if the electric field is oriented in such a way that it
neglects the DMI the Skyrmion gets annihilated. Without DMI the
ferromagnetic configuration is the ground state and the magnetic field
which has stabilized the Skyrmion before destroys it now. The dynamics here 
is similar to the one described in \cite{siemensNJP16ETAL}. Two
phases can be observed: first the reduction of the Skyrmion size and
then the reversal of the center of the Skyrmion. During this
reversal process the Skyrmion releases energy which can be seen in
form of a concentric shock wave running through the system. During the first
phase (reduction of the radius) the Skyrmion shows a twist of the
magnetic moments which ends with the annihilation of the Skyrmion. The
creation process of a Skyrmion is a little bit more complex. Here, the
electric field strengthens the DMI. However, this is not enough to
create the Skyrmion. To create the Skyrmion and the corresponding
topology first the symmetry of the ferromagnetic order needs to be
broken. In other words first a singularity which becomes 
the center of the Skyrmion has to be created. Here, temperature
fluctuations and especially the Joule heating helps. Koshibae and
Nagaosa have shown that a locally increased temperature can be used to
create a Skyrmion \cite{koshibaeNatComm14}. Within the simulations a
local increased temperature $\Delta T = 7.5$ K has been used to
locally break the ferromagnetic symmetry. Another ingredient which can
help to create the Skyrmion is the fact that electric field not only
changes the strength of the DMI, but also the strength of uniaxial
anisotropies  \cite{zhuSciRep14,hibinoAPEX15}. To move the Skyrmion an
electric field with the same field direction as used for the creation
of the skyrmion (in the case of Hsu et al. \cite{hsuARXIV16} pointing away
from the film plane) can be used. Here, the electric field increases
the DMI and therefore decreases the energy of the Skyrmion. The
movement of the STM tip with adequate velocity let the Skyrmion
follow. Thereby the dynamics is the same as using a spin polarized
tunnel current.   

In summary: computer simulations of a Skyrmion in a atomic spin system
with triangular lattice have been performed. It has been shown that it
is possible to use a scanning tunneling microscope to manipulate:
create, annihilate and move a single Skyrmion, either by a spin
polarized current or an electric field. The creation and
annihilation of the Skyrmion reproduces the results of the experiments.
However, the time average of a conventional STM is to low to see the
dynamics during these events. The performed simulations deliver the
missing information and give an explanation for the underlying
physics. The aimed motion of the Skyrmion has not been demonstrated in
experiments so far but the performed simulations show the possibility.


\bibliography{Cite}

\end{document}